\documentclass[10pt,twocolumn,letterpaper]{article}

\usepackage[pagenumbers]{cvpr} 

\usepackage{graphicx}
\usepackage{booktabs}
\usepackage{multirow}
\usepackage{subcaption}

\usepackage{tikz}
\usetikzlibrary{positioning,arrows.meta,calc}

\usepackage{placeins}
\usepackage{listings}
\usepackage{xcolor}
\lstdefinestyle{py}{
  language=Python,
  basicstyle=\ttfamily\small,
  numbers=left,
  numbersep=6pt,
  columns=fullflexible,
  keepspaces=true,
  breaklines=true,
  breakatwhitespace=false,
  postbreak=\mbox{\textcolor{gray}{$\hookrightarrow$}\space},
  frame=tb,
  tabsize=2,
}

\definecolor{cvprblue}{rgb}{0.21,0.49,0.74}
\usepackage[pagebackref,breaklinks,colorlinks,allcolors=cvprblue]{hyperref}

\def\paperID{14392} 
\def\confName{CVPR}
\def\confYear{2026}

\title{CADEvolve: Creating Realistic CAD via Program Evolution}

\author{Maksim Elistratov\textsuperscript{1}\\
\and
Marina Barannikov\textsuperscript{2}\\
\and
Gregory Ivanov\textsuperscript{1}\\
\and
Valentin Khrulkov\textsuperscript{4}\\
\and
Anton Konushin\textsuperscript{1}\\
\and
Andrey Kuznetsov\textsuperscript{3}\textsuperscript{4}\\
\and
Dmitrii Zhemchuzhnikov\textsuperscript{1}\dag
\and \\ \textsuperscript{1}Lomonosov Moscow State University; \\ \textsuperscript{2}Université Paris Dauphine; \\
\textsuperscript{3} Innopolis University; \\
\textsuperscript{4} FusionBrain Lab, AXXX
}

\begin{document}
\maketitle
\let\thefootnote\relax\footnotetext{\textsuperscript{\dag}Corresponding author: zhemchuzhnikovds@my.msu.ru}
\begin{abstract}
Computer-Aided Design (CAD) delivers rapid, editable modeling for engineering and manufacturing. Recent AI progress now makes full automation feasible for various CAD tasks. However, progress is bottlenecked by data: public corpora mostly contain sketch–extrude sequences, lack complex operations, multi-operation composition and design intent, and thus hinder effective fine-tuning. Attempts to bypass this with frozen VLMs often yield simple or invalid programs due to limited 3D grounding in current foundation models. We present CADEvolve, an evolution-based pipeline and dataset that starts from simple primitives and, via VLM-guided edits and validations, incrementally grows CAD programs toward industrial-grade complexity. The result is $\approx$8k complex parts expressed as executable CadQuery parametric generators. After multi-stage post-processing and augmentation, we obtain a unified dataset of $\approx$1.3m scripts paired with rendered geometry and exercising the full CadQuery operation set. A VLM fine-tuned on CADEvolve achieves state-of-the-art results on the Image2CAD task across the DeepCAD, Fusion 360, and MCB benchmarks. Code, dataset, and the SOTA model are available at
\href{https://github.com/zhemdi/CADEvolve}{GitHub},
\href{https://huggingface.co/datasets/kulibinai/cadevolve}{Hugging Face dataset},
and \href{https://huggingface.co/kulibinai/cadevolve-rl1}{Hugging Face model}.
\end{abstract}    
\section{Introduction}

\begin{figure}[t]
  \centering
  \includegraphics[width=\columnwidth]{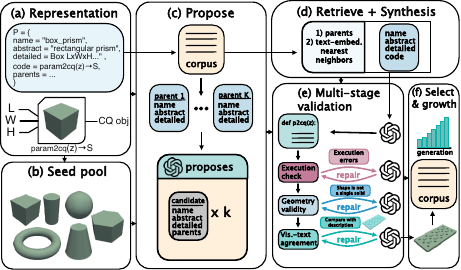}
  \caption{\textbf{CADEvolve overview.}
  (a) Representation of a shape tuple; (b) seed pool of 46 hand-written generators; (c) VLM proposals conditioned on sampled parents; (d) retrieval-augmented code synthesis; (e) staged validation (execution check, geometry validity, visual–text agreement) with targeted repair; (f) selection and growth of the accepted pool.}
  \label{fig:cadevolve-method}
\end{figure}

Computer-Aided Design (CAD) has transformed engineering by enabling precise, parametric modeling and rapid iteration, yet the next leap---AI automation---remains data-limited. In practice, programs are built as sequences of 2D sketches and 3D operations~\cite{willis2021fusion360,wu2021deepcad}; prior corpora and systems represent these either as command tokens~\cite{khan2024cad-signet,wu2021deepcad,chen2025cadcrafter} or as concise, executable Python (e.g., \textsc{CadQuery}~\cite{cadquery})~\cite{rukhovich2024cad-recode,doris2025cad-coder,wang2025gaco}. However, public CAD  {sequence} corpora effectively collapse to sketch--extrude pipelines---e.g., Fusion~360 Gallery, DeepCAD, and CAD-Recode---while richer operations (revolve, loft, sweep, fillet, chamfer, shell, local patterns) are absent from released program histories. This limits the learning of multi-operation composition and design intent. While Seek-CAD~\cite{li2025seek} and RLCAD~\cite{yin2025rlcad} report support for additional operations (e.g., revolve, chamfer , fillet), neither releases multi-operation construction histories, so the community still lacks an open corpus that systematically exercises a broad operator set.

On the geometry side, there are many datasets of real CAD shapes---ABC~\cite{koch2019abc}, ShapeNet~\cite{chang2015shapenet}, MCB~\cite{sangpil2020large}---but they do not provide CAD  {sequences}. DeepCAD~\cite{wu2021deepcad} and Fusion~360 Gallery~\cite{willis2021fusion360} contain parts with user-authored histories, yet their usable sequence splits are predominantly prismatic and sketch--extrude only. To scale data volume, CAD-Recode~\cite{rukhovich2024cad-recode} introduced a rule-based generator, but it still produces only sketches and extrusions; extending it to other operations would require brittle constraint systems to avoid geometric collisions and still does not guarantee richer topology or design intent. A parallel line of work uses  {frozen} VLMs to synthesize code from prompts---e.g., 3D-PreMise~\cite{yuan20243d}, CADCodeVerify~\cite{alrashedygenerating}, and Seek-CAD~\cite{li2025seek}---but single-pass prompting typically yields simple shapes with a narrow operator set, even with retrieval augmentation or multi-stage validation. 

Early in this project, we found that single-pass VLMs struggle to reconstruct  {industrial-grade} CAD programs: they tend to saturate on extruded prisms and fail to chain heterogeneous operations reliably. Recent work suggests a way around this: pair an LLM that proposes code with  {automated evaluators} and  {evolve} candidates via selection, yielding results beyond one-shot capabilities---e.g., the AlphaEvolve coding agent reports new state-of-the-art algorithmic solutions~\cite{novikov2025alphaevolve}. Motivated by this evidence, we introduce \textbf{CADEvolve}: an evolutionary  {propose--execute--filter} pipeline for CAD data generation. Starting from 46 hand-written \textsc{CadQuery} primitives, a VLM (GPT-5-mini in our experiments~\cite{openai2025gpt5}) repeatedly edits and extends parent programs; each candidate must  {compile} to a solid and pass  {geometry checks}. Successful programs become parents for the next round. This pipeline yields \textbf{7{,}945} valid parametric  {generators} (maps from shape parameters $\rightarrow$ \textsc{CadQuery} solids), i.e., each generator represents a class of parts. Unlike EvoCAD~\cite{tobias2025evocad}, which applies evolutionary search at  {inference} time, we use evolution  {only} as an offline data generator.

\begin{figure*}[t]
  \centering
  \includegraphics[width=\textwidth]{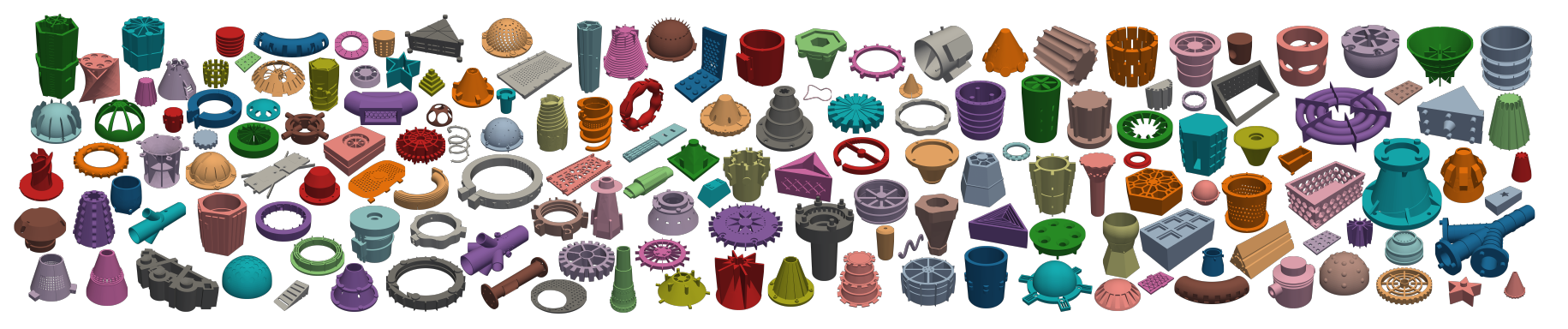}
  \caption{\textbf{Examples of generated parts.}
  A diverse gallery of accepted CADEvolve-G outputs spanning extrude, revolve, loft, sweep, shell, fillet, chamfer, booleans, and local patterns. Colors are arbitrary.}
  \label{fig:generated-parts}
\end{figure*}

From these generators we next build a training corpus. First, we parse each generator and  {sample} parameters, producing $\sim 8{\times}10^5$ runnable programs with paired geometry. Supervised fine-tuning on this set produces diverse, valid multi-operation code, but geometric fidelity to targets remains insufficient. To strengthen supervision  {without altering external datasets}, we run the imperfect model on  {mesh-only} corpora (ABC, ShapeNet) and collect its predicted programs and corresponding shapes as augmentation; we do not modify ABC and ShapeNet themselves---only harvest our model’s outputs on their meshes. We then introduce  {canonicalization}: unify code templates and symbol names, normalize pose and scale, and binarize numeric parameters. This produces $\sim 3.5{\times}10^5$  {canonicalized} scripts after filtering. Finally, we diversify early-tree structure by mixing a balanced, canonicalized subset derived from CAD-Recode primitives with our canonicalized scripts, yielding a $\sim 7{\times}10^5$ training set. Models trained on this set improve substantially;  {RL fine-tuning} with geometry-derived rewards closes the remaining gap and achieves state-of-the-art Image2CAD performance on DeepCAD, Fusion~360 Gallery, and MCB.

For clarity of nomenclature, we denote the three tiers as \textbf{CADEvolve-3L}: 
\textbf{CADEvolve-G} (7{,}945 parametric generators), 
\textbf{CADEvolve-P} ($\sim 8{\times}10^5$ executable programs sampled from generators, with paired geometry), and 
\textbf{CADEvolve-C} ($\sim 8{\times}10^5$  canonicalized normalized binarized scripts used for training). 
Unless stated otherwise, all models are trained on \textbf{CADEvolve-C}. Crucially, CADEvolve uses evolution  {only} for \textbf{offline dataset synthesis}; inference is a single-model decode.

\paragraph{Contributions.}
\begin{itemize}
    \item \textbf{CADEvolve (pipeline).} An offline propose--execute--filter \textbf{evolutionary} pipeline that generates complex multi-operation \textsc{CadQuery} programs, enabling \textbf{realistic synthetic data} when open corpora are small.
    \item \textbf{CADEvolve-3L (dataset).} A three-tier corpus---\textbf{G} (parametric generators), \textbf{P} (executable \textsc{CadQuery} scripts with arbitrary code style), and \textbf{C} (canonicalized programs for training)---that is the \textbf{first open CAD sequence dataset} covering the \textbf{full} \textsc{CadQuery} operator set with executable multi-operation histories.
    \item \textbf{CADEvolve-M (policy).} A vision–language model fine-tuned on \textbf{CADEvolve-C} for the Image2CAD task. It supports the full \textsc{CadQuery} operator set and achieves state-of-the-art reconstruction performance on DeepCAD, Fusion~360 Gallery, and MCB under a fixed architecture and RL recipe.
\end{itemize}
\section{Related Work}

\paragraph{CAD generation.}
Research on CAD synthesis spans three target representations: CSG trees, B-reps, and program/sequence models. CSG focuses on primitive boolean composition and struggles to capture the variety and detail of engineered parts~\cite{du2018inversecsg,ellis2019write,friedrich2019optimizing,kania2020ucsg-net,nandi2018functional,ren2021csg-stump,tian2019learning,yu2022capri-net,yu2023d2csg}. B-rep generators reason over faces, edges and topology but tend to be brittle and harder to edit~\cite{guo2022complexgen,jayaraman2022solidgen,lambourne2021brepnet,li2019supervised,liu2024split,liu2024point2cad,sharma2020parsenet,wang2022neural,wang2020pie-net,xu2024brepgen,li2025caddreamer}. Sequence models --- either command tokens (sketch / extrude / boolean) or concise, executable Python (e.g., \textsc{CadQuery}) --- best match parametric workflows and preserve editability~\cite{wu2021deepcad,lambourne2022prismcad,ren2022extrudenet,xu2022skexgen,xu2023hnc-cad,zhang2024flexcad,badagabettu2024query2cad,chen2024img2cad,khan2024cad-signet,khan2024text2cad,ma2024cad-diffuser,mallis2024cad-assistant,li2025cad-llama,doris2025cad-coder,he2025cad-coder,yuan2025cad-editor,wang2025cad-gpt}. 
Yet the open   {sequence} corpora the community trains on are largely sketch--extrude: Fusion~360 Gallery and DeepCAD provide mostly prismatic histories, while CAD-Recode scales volume via a rule-based extrusion generator~\cite{willis2021fusion360,wu2021deepcad,rukhovich2024cad-recode}. Mesh-only sets (e.g., ABC, ShapeNet, MCB) supply geometry but no editable histories, limiting program-level supervision~\cite{koch2019abc,chang2015shapenet,sangpil2020large}. CADEvolve targets this gap by releasing multi-operation, executable histories.

\paragraph{Case-based CAD program generation.}
To inject domain priors at inference, case-based methods retrieve related designs and structure prompts. Seek-CAD applies RAG and a self-refinement loop for parametric modeling~\cite{li2025seek}. In parallel, several works standardize code formats to improve executability and editing --- OpenECAD, InstructGraph --- and CAD-Llama’s SPCC adds a hierarchical semantic layout~\cite{zhe2024openecad,jianing2024instructgraph,li2025cad-llama}. These strategies curb hallucinations but ultimately inherit the limits of available corpora: while Seek-CAD and RLCAD report support for revolve, fillet and chamfer, corresponding multi-operation histories are not publicly released~\cite{li2025seek,yin2025rlcad}. CADEvolve provides such histories explicitly, together with a canonicalized layer tuned for training.

\paragraph{Evolutionary methods.}
LLM-driven evolution --- propose, evaluate, and select --- has enabled problems beyond single-pass generation: FunSearch, LLaMEA / LLaMEA-HPO, and AlphaEvolve demonstrate competitive algorithm discovery via tight validator loops~\cite{bernardino2024funsearch,stein2024llamea,stein2024llameahpo,novikov2025alphaevolve}. In CAD, EvoCAD deploys evolution   {at inference}, maintaining a population ranked by text similarity and refined via crossover and mutation with a self-debug filter~\cite{tobias2025evocad}. 
CADEvolve relocates evolution to the   {data stage}: starting from hand-written \textsc{CadQuery} primitives, a VLM proposes edits; only programs that compile to solids and pass geometry checks survive. This offline   {propose--execute--filter} process yields parametric   {generators} and large batches of executable multi-operation scripts suitable for pretraining; no population search is used at test time.

\paragraph{Image2CAD.}
Recent work reconstructs CAD sequences directly from visual inputs. Early single-view approaches such as CSGNet predict compact CSG programs from a single raster image of a synthetic shape~\cite{sharma2018csgnet}. Subsequent work has framed Image2CAD as a reinforcement-learning problem, optimizing token-level policies against rendering- or geometry-based rewards and scaling to richer CAD corpora~\cite{chao2025reinforcement,yin2025rlcad}. More recent models --- CADCrafter, CADCoder, and  \emph{cadrille } --- consume multi-view grids or isometric renderings of CAD parts and emit parametric programs that better match industrial design workflows~\cite{chen2025cadcrafter,doris2025cad-coder,kolodiazhnyi2025cadrille}. In this work, we adopt Image2CAD as the simplest controlled setting to validate datasets: fixed multi-view renderings fully specify the target geometry without requiring per-shape textual descriptions (as in Text2CAD) or additional point-cloud encoders (as in PC2CAD), making architectural confounders minimal and comparisons more transparent.

\paragraph{RL for CAD reconstruction.}
Two lines exist. (i) Environment-level RL optimizes command sequences with geometry-based rewards without LLMs (REINFORCE in CSGNet; DQN for orthographic drawings; RLCAD’s B-rep gym)~\cite{sharma2018csgnet,chao2025reinforcement,yin2025rlcad}. (ii) LLM post-training aligns code generators using verifiable signals: DPO with code-checkers or visual feedback (CADCrafter, CADFusion) and GRPO-family objectives with geometry-aware rewards (CAD-Coder, \textit{cadrille}, GACO-CAD)~\cite{chen2025cadcrafter,wang2025cadfusion,guan2025cad-coder-grpo,kolodiazhnyi2025cadrille,wang2025gaco}. 
Our contributions are orthogonal to these RL recipes: we re-use standard GRPO-style objectives and baselines, but introduce a  {multi-operation, executable, canonicalized} corpus that benefits both SFT-only and RLVR pipelines, providing the operator diversity and code regularity missing from prior open datasets.
\section{Dataset Generation and Processing}

\subsection{Evolutionary Synthesis of Parametric Generators (CADEvolve-G)}
\label{subsec:cadevolve_g}
\begin{figure}[b]
  \centering
  \includegraphics[width=\columnwidth]{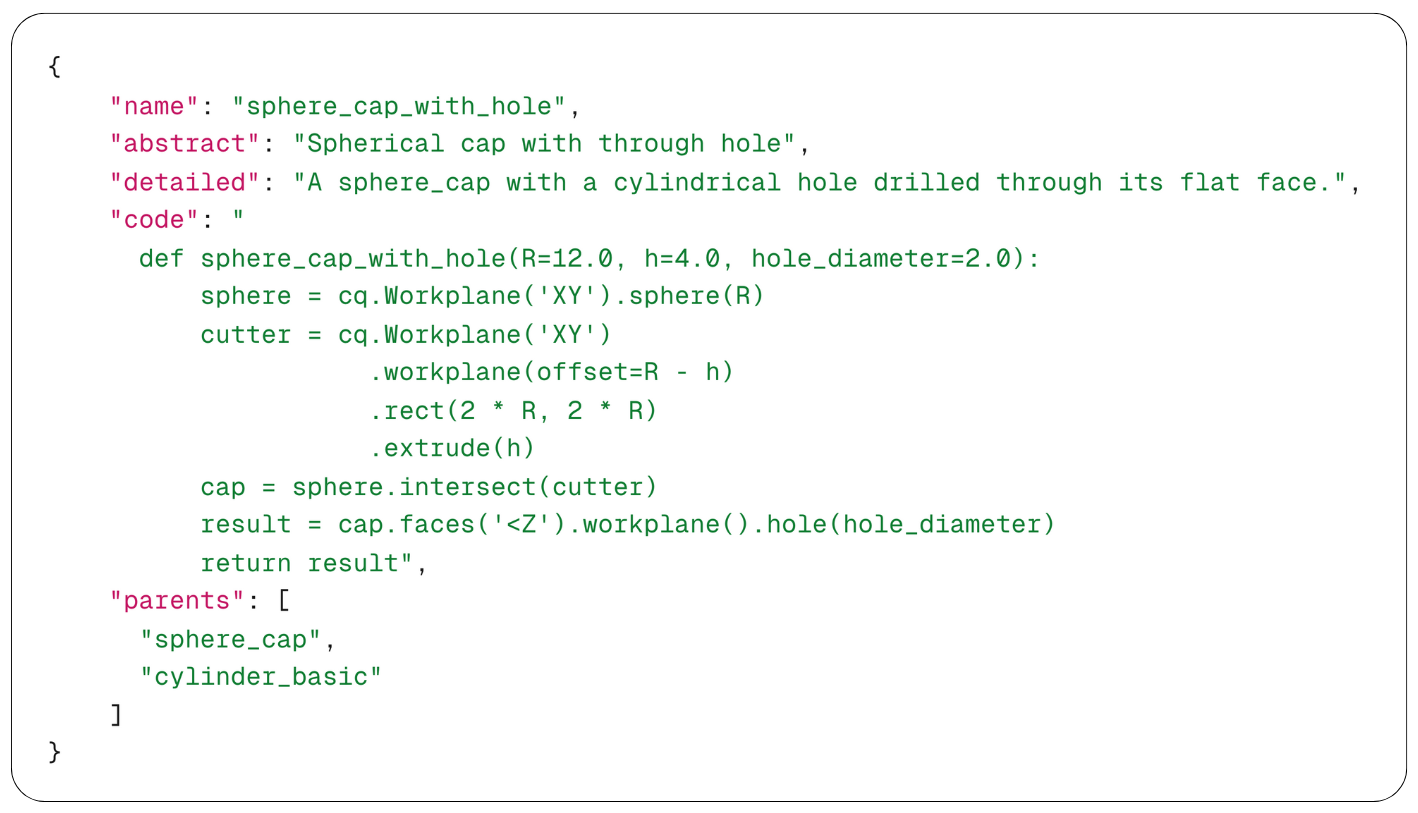}
  \caption{\textbf{Representation of a shape from CADEvolve-G.}
  The representation consists of descriptive textual fields, a Python code that maps geometric parameters to a 3D shape \textit{S}, and the list of parents from which \textit{S} has evolved via the CADEvolve algorithm.}
  \label{fig:shaperepr}
\end{figure}

\paragraph{Representation.}
We represent a shape as a tuple
$\mathcal{P}=\{\textit{name},\,\textit{abstract},\,\textit{detailed},\,\textit{code},\,\textit{parents}\}$
(Fig.~\ref{fig:shaperepr}),
where $\textit{name}$ specifies the part name in the  snake case,   $\textit{abstract}$ and $\textit{detailed}$ are abstract (concise description) and detailed descriptions, respectively, the code is the mapping \mbox{$\textit{param2cq}:\mathbf{z}\!\mapsto\! S$}, a self--contained \textsc{CadQuery} function mapping a collection of semantic parameters $\mathbf{z}$
to a single watertight solid $S$. For different values of parameters, we obtain different solids, e.g., gears of varying radii.
The textual fields capture design intent; \textit{parents} records lineage for inheritance.

\paragraph{Seed pool.}
We begin with an initial corpus of \textbf{46} hand-written generators that collectively cover \texttt{extrude}, \texttt{revolve}, \texttt{loft}, \texttt{sweep}, \texttt{shell}, \texttt{fillet}, \texttt{chamfer}, booleans, and local patterns/arrays (e.g., gears, wedges, prisms, torus segments).
This pool anchors operator breadth and parameterization styles.

\paragraph{Propose–Execute–Filter loop.}
The initial pool $\mathcal{D}_0$ is seeded by the previously discussed manually constructed programs. Given the current accepted pool $\mathcal{D}_t$ we perform the following steps:
\begin{enumerate}
  \item \textbf{Parent sampling.} Randomly sample $K$ parents from $\mathcal{D}_t$ to encourage recombination across operations.
  \item \textbf{Child metadata proposal.} \emph{gpt5-mini} is asked to propose  $k$ children, each providing
  \emph{name}, \emph{abstract}, \emph{detailed},
  and the list of \emph{parents} it inherits from. Proposals must imply a single solid body and 
  avoid an assembly of repeating solids and encourage more complex geometry than their parents.
  \item \textbf{Code synthesis with retrieval.} For each child, we retrieve a small set of nearest neighbors
  by embedding the {detailed} description and union it with the sampled parents’ code. \emph{gpt5-mini} then produces a
  {monolithic, parametric} \textsc{CadQuery} function $\textit{param2cq}$, explicitly exposing design--meaningful parameters with default values that are used to call the generator and valid the execution and the output shape.
  \item \textbf{Staged validation and self-repair.}
    \begin{itemize}
      \item Execution check: $\textit{param2cq}$ must compile and run on defaults, returning exactly one solid.
      \item Geometry validity: the result passes strict CAD integrity tests.
      \item Visual–text agreement: we render a seven-view montage (one isometric + six orthographic projections) and ask the VLM to verify that the rendered geometry matches the child’s abstract and detailed descriptions. If any stage fails,
      the model is prompted to issue a targeted fix.
    \end{itemize}
  \item \textbf{Selection and growth.} Only children that pass all verification stages are admitted to $\mathcal{D}_{t+1}$; we store their metadata and lineage. The loop repeats until a fixed budget is met or novelty saturates (see Supplementary~\hyperref[sec:novelty_dynamics]{A}).
\end{enumerate}
This procedure yields \textbf{7{,}945} validated parametric generators CADEvolve-G. An overview of the pipeline is shown in Fig.~\ref{fig:cadevolve-method}, and representative evolutionary lineages are visualized in Supplementary~\hyperref[sec:suppl_evo_traj]{B}.



\subsection{Sampling and Parsing of Generators (CADEvolve-P)}
\label{sec:sampling-parsing}


\paragraph{Goal.}
Given this set of generators, we extend it as follows. For each parametric generator, we extract its set of parameters  with default values $\mathbf{z}$ and search for a small diverse set of variations of this initial vector
${\mathbf{z}_1,\ldots,\mathbf{z}_N}$ (we use $N\!=\!15$) that (i) produce valid solids and (ii) cover distinct regions of the design space $\textit{param2cq}$.

\paragraph{Quality–diversity objective.}
We define our fitness objective as follows. Given an arbitrary parameter vector as an input to the generator, we build the corresponding shape and compute the penalty with two terms.
(1) \emph{Validity/fit:} If CAD checks fail (not exactly one watertight solid), assign a large penalty. Otherwise, add small penalties if the longest side falls outside the $[60,200]$ unit range or if any face of the axis-aligned bounding box exits the cube $[-100,100]^3$.
\\
(2) \emph{Novelty:} Compare the candidate to an archive of accepted samples. If it lies closer than a distance threshold $\varepsilon$ to any archived point, add a non-negative penalty that grows as the gap to $\varepsilon$ increases. When both parts are zero, the sample is valid and novel.

\paragraph{Search.}
Given the non-differentiable nature of the task, we use the well-known black box optimization approach CMA-ES ~\cite{hansen2023cmaevolutionstrategytutorial}  with default parameters as the initial vector 
to find unique parameter vectors producing valid shapes.
We iterate until we collect N accepted samples per generator (we use $N{=}15$) or a compute budget is reached, yielding a compact set of valid, diverse instances for each generator.

\paragraph{From generators to concrete programs.}
For each accepted parameter vector $\mathbf{z}$ and its generator $\textit{param2cq}$, we emit a deterministic \textsc{CadQuery} script, termed CADEvolve-P, via single-run tracing and slicing:
\begin{enumerate}
\item \textbf{Header.} Insert minimal imports (\texttt{import cadquery as cq}; \texttt{import math} only if used).
\item \textbf{Parameter materialization.} Emit one bind per argument in the exact order expected by $\textit{param2cq}$: \texttt{param\_i = z["param\_i"]}.
\item \textbf{Trace \& slice.} Execute $\textit{param2cq}(\mathbf{z})$ under a tracer; record only constructive \textsc{CadQuery} operations that contribute to the final solid (sketch ops; extrude/revolve/loft/sweep; fillet/chamfer/shell; booleans; patterns). Drop checks, logging, and other non-geometric statements.
\item \textbf{Control-flow resolution.} Keep only executed branches and realized loop bodies; no residual \texttt{if}/\texttt{try} remain in the output.
\item \textbf{Standardize the output variable name.} Emit a flat, side-effect–free sequence with stable formatting, minimal imports, and a unified tail line \texttt{result = \ldots}.
\end{enumerate}

\begin{figure}[t]
  \centering
  \includegraphics[width=\columnwidth]{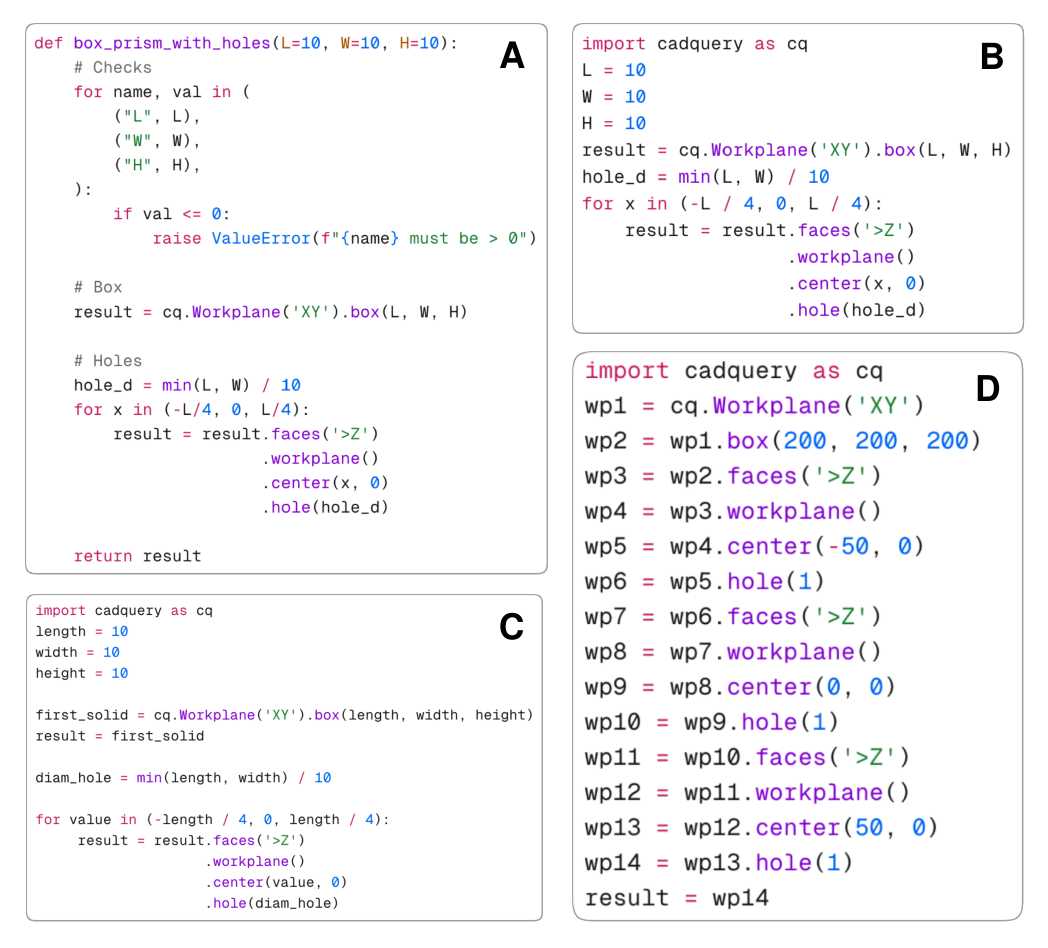}
  \caption{\textbf{From generator to concrete program.}
  Given a (A) parametric generator $\textit{param2cq}$ and a sampled parameter vector $\mathbf{z}$, we (B) bind parameters (\texttt{param\_i = z["param\_i"]}), then execute the generator once to resolve conditionals and loops, skipping untaken branches and retaining only geometry-affecting \textsc{CadQuery} operations; (C) apply code–level augmentation; and (D) emit a flat, deterministic script with minimal imports and a unified output
  (\texttt{result = \dots}). The script reproduces the shape exactly while exposing its construction history.}
  \label{fig:gen2script}
\end{figure}

The resulting script contains parameter binds from $\mathbf{z}$; it deterministically reproduces the shape and makes the construction history explicit. These scripts form CADEvolve-P and are used downstream for training and evaluation (Fig.~\ref{fig:gen2script}). We target 10 scripts per generator; this stage yields \mbox{74{,}918} scripts in total.

\subsection{Direct Image2CAD training and code augmentation}

Starting from $\approx\!75$k generator–script pairs, we observed severe template collapse: within each generator, scripts shared the same identifiers and operation sequence, differing only in argument values. A small Qwen2-VL-2B model trained on this data learned spurious correlations between geometry and a fixed code skeleton and reproduced training templates; results were unusable, so we omit metrics. To break this bias while preserving geometry, we applied code--level augmentation: for every script we asked a compact LLM (gpt-5-mini) to produce up to 10 semantically equivalent rewrites (different structure, same solid). We over-generated and kept only validated, executable variants, yielding in total \textbf{744{,}780} scripts. During this pass we also pruned non-contributing operations (e.g., dead sketches, unused workplanes, no-op fillets), keeping each script minimal yet functionally identical.

\begin{table*}[t]
\centering
\caption{\textbf{CAD sequence generation conditioned on multi-view images.}
When CADEvolve-M (RL1) is trained on the same RL set as cadrille , it achieves lower CD and higher IoU than cadrille on all three datasets, at the cost of higher IR, reflecting more frequent use of complex operations that are more collision-prone than sketch–extrude pipelines.
Adding MCB to the RL training pool  trades a small drop in CD/IoU on DeepCAD and Fusion360—bringing them close to cadrille’s levels—for a substantial improvement on MCB.}
\label{tab:img_results}
\setlength{\tabcolsep}{7pt}
\begin{tabular}{lccc ccc ccc}
\toprule
& \multicolumn{3}{c}{DeepCAD} & \multicolumn{3}{c}{Fusion360} & \multicolumn{3}{c}{MCB} \\
\cmidrule(lr){2-4}\cmidrule(lr){5-7}\cmidrule(lr){8-10}
Method & CD$\downarrow$ & IoU$\uparrow$ & IR$\downarrow$ & CD$\downarrow$ & IoU$\uparrow$ & IR$\downarrow$ & CD$\downarrow$ & IoU$\uparrow$ & IR$\downarrow$ \\
\midrule
\emph{cadrille} SFT                 & 0.19 & 86.5 & 1.6 & 0.20 & 77.3 & 3.4 & 1.16 & 40.4 & 14.3 \\
\emph{cadrille} RL                  & 0.17 & 92.2 & \textbf{0.1} & 0.17 & 84.6 & \textbf{0.1} & 0.87 & 47.6 & 2.5 \\
\midrule
 {CADEvolve-P pre-aug} (SFT)    & 7.31 & 37.2 & 17.3 & 9.15 & 29.7 & 19.1 & 13.19 & 17.3 & 25.3 \\
 {CADEvolve-P post-aug} (SFT)   & 4.93 & 42.9 & 14.3 & 7.54 & 33.1 & 16.1 & 10.98 & 20.2 & 22.3 \\
\midrule
 {CADEvolve-C small} (SFT)      & 3.40 & 49.6 & 19.1 & 9.25 & 35.1 & 26.1 & 8.45 & 28.1 & 43.1 \\
 {CADEvolve-C middle} (SFT)     & 0.57 & 70.1 & 23.6 & 0.68 & 59.1 & 25.2 & 2.09 & 39.1 & 37.7 \\
 {CADEvolve-C big} (SFT)        & 0.67 & 72.1 & 19.0 & 0.26 & 71.1 & 18.2 & 1.71 & 42.0 & 32.0 \\
\midrule
 {CADEvolve-C big} (RL1)        & \textbf{0.15} & \textbf{92.6} & \textbf{0.2} & \textbf{0.16} & \textbf{87.2} & 0.5 & 0.62 & 51.4 & 2.3 \\
 {CADEvolve-C big} (RL2)        & 0.16 & 91.1 & 0.1 & \textbf{0.16} & 84.0 & 0.2 & \textbf{0.52} & \textbf{55.2} & \textbf{0.4} \\
\bottomrule
\end{tabular}
\end{table*}

\subsection{Bootstrapping with Image2CAD pretraining and mesh2CAD distillation}

\paragraph{Round 1 (Image2CAD on rewrites).}
We trained Qwen2-VL-2B on Image2CAD using the \textbf{744{,}780} rewritten scripts as targets (multi-view renders as inputs). The model produced diverse, valid code but shapes were far from the targets; quantitative results are deferred to Sec.~\ref{sec:experiments}.

\paragraph{Round 2 (adding ABC/ShapeNet predictions).}
To expand coverage, we used the Round-1 model to predict \textsc{CadQuery} programs for meshes from ABC and ShapeNet, then filtered to keep only valid scripts whose bounding-box max-extent lies in $[60,200]$. This produced \textbf{875{,}632} ABC scripts and \textbf{119{,}437} ShapeNet scripts. Combined with our rewrites, the training set totaled \textbf{1{,}739{,}849} scripts ($\approx\!1.74$M). We retrained Qwen2-VL-2B on the enlarged corpus; metric gains were incremental and still below state-of-the-art, motivating further dataset improvements (Sec.~\ref{sec:experiments}).

\subsection{CADEvolve-C: Canonicalization \& Normalization}

Our initial corpus, while valid, was overly sophisticated and exhibited wide scale variation (max extent in $[60,200]$). We therefore enforce a unified format, size, and numeric grid so the learner focuses on construction logic rather than incidental syntax or scale.

\begin{enumerate}
\item \textbf{Unification.} Remove residual non-geometric Python; keep only geometry-affecting \textsc{CadQuery} calls. Re-emit as a flat, macro-like sequence with stable temporaries (\texttt{wp1}, \texttt{wp2}, \dots) and minimal imports.
\item \textbf{Centering.} Build the solid, compute its AABB center, and inject a deterministic translation so the final object is centered at $(0,0,0)$.
\item \textbf{Extent normalization.} Apply a uniform scale so the bounding box longest side equals a fixed target (200 units), yielding a canonical size (roughly within $[-100,100]^3$).
\item \textbf{Binarization.} Quantize all numeric literals after scaling: zero-out tiny epsilons and round remaining values to integers. This removes floating-point noise and constrains the parameter space to a consistent grid.
\end{enumerate}

As illustrated in Fig.~\ref{fig:gen2script}, we convert traced generators into canonical, centered, and uniformly scaled \textsc{CadQuery} programs before training.

\paragraph{Collision-aware pruning after CADEvolve-C.}
\label{sec:finalset}
Canonicalization occasionally induced geometric collisions. We re-validated all scripts post-transform and kept only valid ones, yielding 
\textbf{1{,}002{,}002} programs in total: \textbf{69{,}201} from CADEvolve, \textbf{813{,}378} ABC predictions, and \textbf{119{,}312} ShapeNet predictions. Note that canonicalization was applied to  CADEvolve-P before code-style augmentation; the rewrite pass mainly enabled diverse ABC/ShapeNet predictions but they are not part of this final canonicalized layer since the output of canonization of different rewrites for one shape is identical. 

\paragraph{Length-based filtering and truncation.}
We split the original dataset into two groups based on script length: \textbf{849{,}558} scripts had fewer than 3k characters, while \textbf{152{,}444} scripts exceeded 3k characters. For the longer scripts, we applied truncation to 3k characters and successfully truncated \textbf{151{,}892} scripts. We then passed these truncated scripts through the canonicalization pipeline again (excluding the unification stage, since it had already been applied). As a result, we obtained \textbf{129{,}961} valid scripts. After a final deduplication step, these corresponded to \textbf{111{,}742} unique scripts. Therefore, the final number of scripts shorter than 3k characters is \textbf{961{,}300}.

\paragraph{Post-canonization baseline.}
Training \textsc{Qwen2-VL-2B} on this set improved results markedly but remained below SOTA (see Sec.~\ref{sec:experiments}). Error analysis pointed to limited sketch diversity.

\paragraph{Sketch-diversity augmentation.}
To inject variability in early sketches, we targeted CADEvolve scripts whose canonicalized first primitives hit the normalization bound (e.g., axis-aligned \texttt{box} with max extent 200 or \texttt{cylinder} with diameter/height 200). We replaced the base primitive with a script from CADRecode (known for diverse sketches). This contributed \textbf{963{,}096} additional scripts. 

\paragraph{Mesh generation.}
Using the scripts from the length-based filtering step and the sketch-diversity augmentation step, we generated STL meshes for the resulting set of scripts. Due to rendering and geometry validation failures, STL files were successfully produced for only \textbf{1{,}382{,}928} scripts.

\paragraph{Rotational augmentation.}
Using rotational augmentation (see Supplementary~\hyperref[sec:rot_aug]{C}), we obtained an additional \textbf{1{,}337{,}553} scripts. The final supervised fine-tuning (SFT) training set thus comprises \textbf{2{,}720{,}481} scripts.

\paragraph{Dataset characteristics.}
Key dataset characteristics, including operation occurrence statistics, sequence length, and geometric face count, are reported in Supplementary~\hyperref[sec:suppl_benchmarks]{D}.

\section{CADEvolve-M: Program-Generating Policy}
\label{sec:experiments}

\subsection{Problem and Training Pipeline}
\label{subsec:problem-pipeline}
\begin{figure*}[t]
  \centering
  \includegraphics[width=\textwidth]{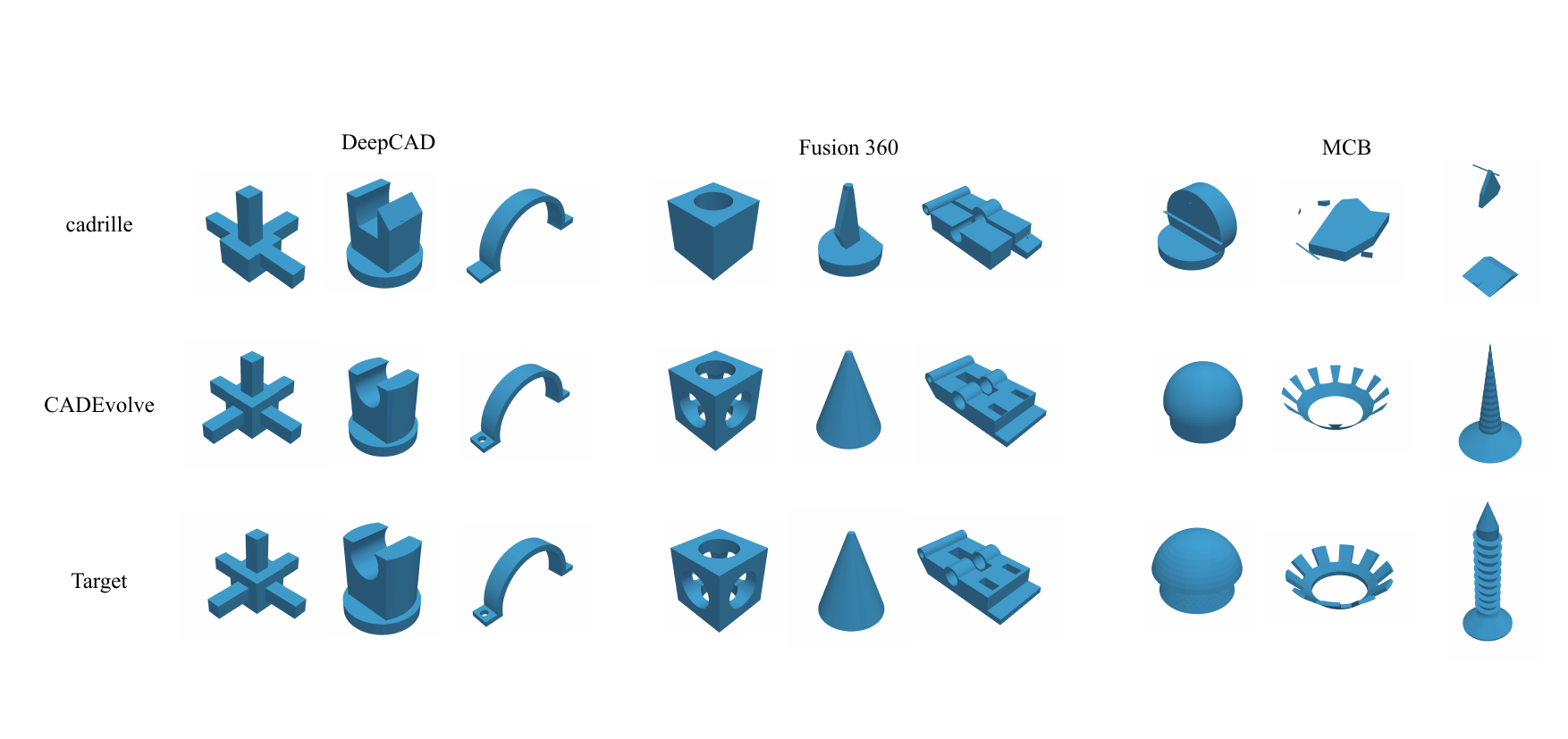}
  \caption{\textbf{Qualitative comparisons across datasets.}
  Columns: \emph{DeepCAD}, \emph{Fusion360}, \emph{MCB}. 
  Rows (top$\to$bottom): the \emph{cadrille} baseline, our  {CADEvolve\textendash C big (RL2)} prediction, and the target render.
  This panel best illustrates the advantages of the  {CADEvolve} dataset: targets include parts built via \emph{revolve},
  \emph{sweep}, \emph{loft}, face selectors, and complex hole patterns that cannot be well approximated by the sketch- extrude-boolean schemes used in CAD-Recode and many other datasets, where \emph{cadrille} typically fails but  
  {CADEvolve-M}
    closely
    reconstructs 
    the input shape.
    }
  \label{fig:qual_cherry}
\end{figure*}

\paragraph{Problem.}
To validate the dataset, we adopt the simplest controllable setting:  {Image2CAD}.
Given a fixed multi-view render 
of a shape, the model emits a \textsc{CadQuery} program that compiles to a solid matching the target.
Unlike  {Text2CAD}, this task does not need textual descriptions, which we do not have for scripts from augmentations; unlike {PC2CAD}, it does not require training a point-cloud encoder.
We feed the multi-view image grid directly into the VLM’s built-in visual encoder, introducing no extra image backbone, adapters, or pretraining beyond the base model.

\begin{figure}[t]
  \centering
  \includegraphics[width=\columnwidth]{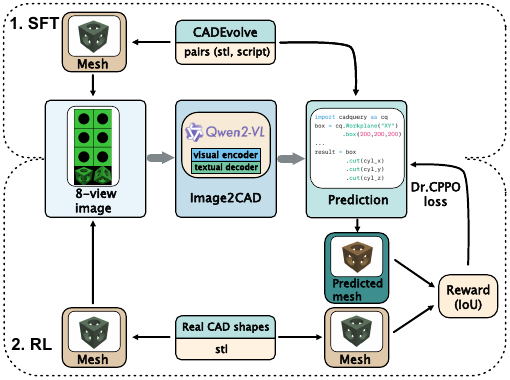}
  \caption{\textbf{Image2CAD training pipeline.}
  Multi-view inputs (7 views for CADEvolve-P; 8 canonical views for CADEvolve-C/RL) are fed to the VLM’s built-in visual encoder; a textual decoder emits \textsc{CadQuery} code. We first run SFT on paired (render, script) data, then apply online RL (Dr.\,GRPO + CPPO) with an IoU-based reward and invalidity penalties, using the predicted mesh for feedback.}
  \label{fig:cadevolve-m}
\end{figure}

\paragraph{Experiment setup.}
Experiments on CADEvolve-P used $7$ views (6 orthographic $+$ 1 iso); final experiments on CADEvolve-C use $8$ canonical views: six orthographic projections ($\pm X,\pm Y,\pm Z$) and two isometric views. Shapes are rigid-aligned and lie in $[-100,100]^3$. For each orthographic view we render a $238{\times}238$ image of that box and encode depth along the view axis via intensity. To keep axis directions consistent, $-Z$, $+Y$, $+X$ images are horizontally mirrored. The eight images are concatenated into a $2{\times}4$ grid fed to the model. Unlike the 4-iso setup in \emph{cadrille} \cite{kolodiazhnyi2025cadrille}, the 6-ortho + 1–2 iso layout sharpens cues for fillets and chamfers. We use \textsc{Qwen2-VL-2B} as the vision–language backbone, prompting it with the multi-view grid and decoding \textsc{CadQuery} tokens. For experiments on CADEvolve-P we normalized the target shape before visualizations and put its center and maximum extent in the prompt which is not needed for CADEvolve-C where all the shapes are normalized. An overview of the end-to-end Image2CAD training pipeline (SFT→RL) is shown in Fig.~\ref{fig:cadevolve-m}.

\subsection{Datasets}
\label{subsec:cadevolve-datasets}
\paragraph{SFT corpus.} We did experiments on different steps of CADEvolve-P and CADEvolve-C processing.

\paragraph{RL \& Evaluation data.}
We ground both RL fine-tuning and evaluation on three public corpora:  {DeepCAD},  {Fusion360}, and  {MCB}. To ensure coverage and reduce category bias, we re-split  {MCB} so that the test set spans all ISO categories. For RL, both runs use the \emph{cadrille} RL train set (selected parts from the DeepCAD and Fusion360 train splits); 
in the second run, we additionally include the MCB training split while keeping its test split fixed

\subsection{Metrics}
\label{subsec:cadevolve-metrics}
All meshes are rigid-aligned and normalized to $[0,1]^3$. We report (i) \textbf{Chamfer Distance (CD)} on $8{,}192$ vs $8{,}192$ points (scaled by $10^3$), (ii) \textbf{volumetric IoU} (\%), and (iii) \textbf{Invalid Rate (IR)} --- fraction of generations that fail to compile or yield a non-watertight/degenerate solid. To reduce invalidity bias we report \emph{median} CD.

\subsection{Training}
\paragraph{SFT.}
We perform two epochs of supervised fine-tuning in each experiment. Objective: token-level cross-entropy on code conditioned on views. 

\paragraph{RL fine-tuning.}
We adopt the same online RL training and reward as \emph{cadrille}, i.e., the GRPO objective with Dr.\,GRPO and CPPO variants (Dr. CPPO) \cite{shao2024deepseekmath, liu2025dr.grpo, lin2025cppo}, and a programmatic reward that combines \emph{IoU} (scaled to emphasize accuracy) with a penalty for invalid generations (non-compiling or non-watertight)~\cite{kolodiazhnyi2025cadrille}. 
The reward is $r = 10 \cdot \mathrm{IoU}$ if code compiles  otherwise $r = -10$.

We train on two configurations for 20 epochs each:
\begin{itemize}
  \item \textbf{RL 1}: RL on the \emph{cadrille}  RL train set (selected parts from DeepCAD and Fusion360 train splits, absent from the SFT corpus).
  \item \textbf{RL 2}: same as RL 1 $+$ MCB train split. MCB is re-split by us so the test set covers all ISO categories and is never used in RL.
\end{itemize}

MCB exhibits a distinct rendering domain: mesh export uses relatively high STL tolerances that smooth sharp edges, producing softer silhouettes than DeepCAD and Fusion360. This alters view-space cues and introduces a domain shift. 
{RL1} reproduces the \emph{cadrille} protocol (RL on DeepCAD+Fusion360 only) for a like-for-like comparison; we expected weaker generalization to MCB under this shift.
{RL2} adds the MCB train split to the RL pool—while keeping our MCB test fixed—to explicitly adapt to this smoother visual regime.

\paragraph{Baseline choice.}
We benchmark primarily against \emph{cadrille} because  it is the state-of-the-art in the image-only setting, offers public code, uses the same RL algorithm (Dr.\,CPPO) and matching evaluation metrics. 
\begin{table}[t]
\centering
\caption{\textbf{Dataset/Regime definitions} used in Table~\ref{tab:img_results}. “Canon.” = canonicalized (centered, scaled, quantized).}
\label{tab:regimes}
\setlength{\tabcolsep}{4pt}
\begin{tabular}{p{2.5cm}lp{3.5cm}}
\toprule
Tag & Sup. & Composition / Notes (view protocol) \\
\midrule
\emph{cadrille} SFT/RL & SFT/RL & Published baseline; RL with Dr.\,CPPO;  {4 isometric} views. \\
\midrule
CADEvolve-P pre-aug & SFT & Traced generator scripts only (CADEvolve-P); no code rewrites; no ABC/ShapeNet; canon. \ off;  {7 views} (6 ortho + 1 iso). \\
CADEvolve-P post-aug & SFT & + Semantics-preserving code rewrites; pruned non-contributing ops; still no ABC/ShapeNet; canon.\ off;  {7 views}. \\
\midrule
CADEvolve-C small  & SFT & Generator-only regime, \emph{canonicalized}; 8 views (6 ortho + 2 iso). \\
CADEvolve-C middle & SFT & + ABC/ShapeNet predictions (\emph{canonicalized}); no CADRecode mix; 8 views. \\
CADEvolve-C big    & SFT & Middle + mix with CADRecode (\emph{canonicalized}); 8 views. \\
\midrule
CADEvolve-C big (RL1) & RL & SFT init (C big) $\rightarrow$ RL on \emph{cadrille} RL train (DeepCAD+Fusion360); 8 views. \\
CADEvolve-C big (RL2) & RL & As RL1 $+$ \emph{MCB train} split; 8 views. \\
\bottomrule
\end{tabular}
\end{table}

\subsection{Results}
\label{subsec:results}

Table~\ref{tab:img_results} summarizes Image2CAD performance on \textsc{DeepCAD}, \textsc{Fusion360}, and \textsc{MCB}. We report \textbf{Median CD}$\downarrow$ ($\times 10^3$), \textbf{mean IoU}$\uparrow$ (\%), and \textbf{IR}$\downarrow$ (invalid rate).  

We compare (i) the \emph{cadrille} baselines (SFT/RL), (ii) \textsc{CADEvolve-P} before/after code-level augmentation, and (iii) \textsc{CADEvolve-C} under progressively stronger data regimes (small/middle/big), followed by RL fine-tuning from the \textsc{CADEvolve-C big} SFT checkpoint (RL1/RL2). 

Even after code-level augmentation, \textsc{CADEvolve-P post-aug (SFT)} remains far from the strongest baselines , although augmentation does move metrics in the right direction compared to pre-augmentation .  
This supports the interpretation that semantics-preserving rewrites reduce template overfitting, but are insufficient without canonization of the code style and shape size.

Moving from \textsc{CADEvolve-C small} to \textsc{middle} yields a large jump , indicating that adding prediction-derived supervision (ABC/ShapeNet) substantially improves geometric fidelity.  The \textsc{big} regime further improves performance.  

Starting from \textsc{CADEvolve-C big (SFT)}, RL fine-tuning produces strong results across all three datasets. In RL1  improves over \emph{cadrille RL} in CD/IoU on all datasets , at the cost of a slightly higher invalid rate, consistent with more frequent use of complex, collision-prone operations. 
RL2 augments the RL pool with \textsc{MCB} training shapes to address its domain shift (softer silhouettes due to higher STL tolerances). 
This yields a substantial improvement on MCB  while maintaining near-\emph{cadrille} performance on DeepCAD and Fusion360 .  

\section{Limitations}
\label{sec:limitations}

While CADEvolve is designed to provide a large-scale, diverse corpus of \emph{validated} parametric CAD programs spanning a broad range of operators and geometric complexity, it is important to acknowledge several limitations of the proposed generation process and resulting dataset.


\textbf{Synthetic distribution mismatch.}
CADEvolve is a synthetic dataset produced by an evolution loop and is not intended to match any single proprietary industrial CAD distribution. Consequently, the induced shape and operation frequencies may differ from real-world data. Despite this, we observe improved generalization across multiple benchmarks including stronger performance on MCB in our experiments, but we do not claim distribution-level fidelity to any particular industrial domain.

\textbf{CadQuery dialect scope.}
The generated programs are expressed in CadQuery. Many operations are conceptually portable (e.g., extrude, revolve, loft, sweep, fillet/chamfer, booleans), but faithful conversion to other CAD systems may be non-trivial due to differences in feature-history representations, kernel behaviors, and constraint semantics.

\section{Conclusion}

We proposed \textbf{CADEvolve}, a general method for synthesizing high-quality supervision when open corpora are scarce, and instantiated it in the CAD domain. The resulting CADEvolve-3L dataset is  the first CAD sequence corpus covering the full CAD operation set, and it already yields state-of-the-art Image2CAD performance, suggesting that the same data foundation can further boost PC2CAD/Scan2CAD, Text2CAD, and broader multimodal CAD pipelines.

\small
\bibliographystyle{ieeenat_fullname}
\bibliography{main}

\appendix
\clearpage
\section{Evolutionary synthesis trajectories}
\label{sec:suppl_evo_traj}

\begin{figure}[h]
\centering
\includegraphics[width=0.7\columnwidth]{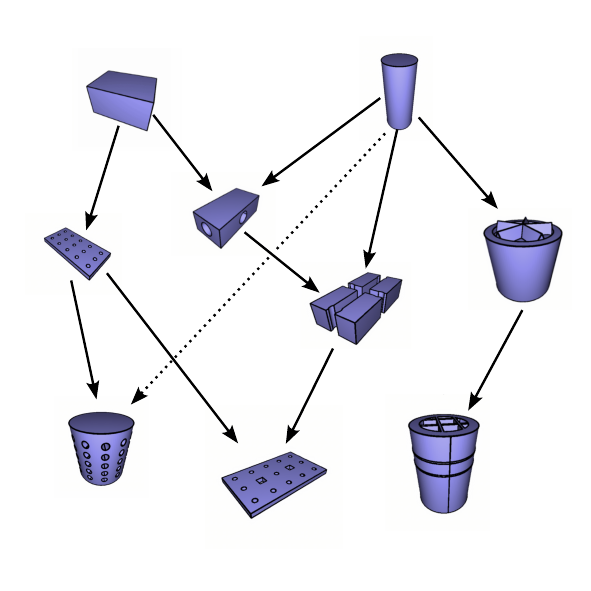} 
\caption{\textbf{Example evolutionary trajectories.} Nodes correspond to generated CAD parts; directed edges indicate parent$\rightarrow$child refinements proposed by the LLM. Over iterations, the shapes become progressively more complex through compositional multi-operation edits.}
\label{fig:evolution}
\end{figure}

Fig.~\ref{fig:evolution} visualizes representative trajectories produced by our evolutionary \emph{propose--execute--filter} pipeline driven by a LLM. Starting from a small set of simple seed primitives (top), the model proposes incremental code edits that introduce new operations and structural detail. Accepted candidates become parents for subsequent iterations, yielding branching lineages and occasional recombination across different design directions. Overall, the graph illustrates how complexity is accumulated progressively — from basic solids to multi-operation, higher-detail parts — under automated execution and validation constraints.

\section{Novelty--validity dynamics}
\label{sec:novelty_dynamics}

\begin{figure}[h]
  \centering
  \includegraphics[width=0.8\columnwidth]{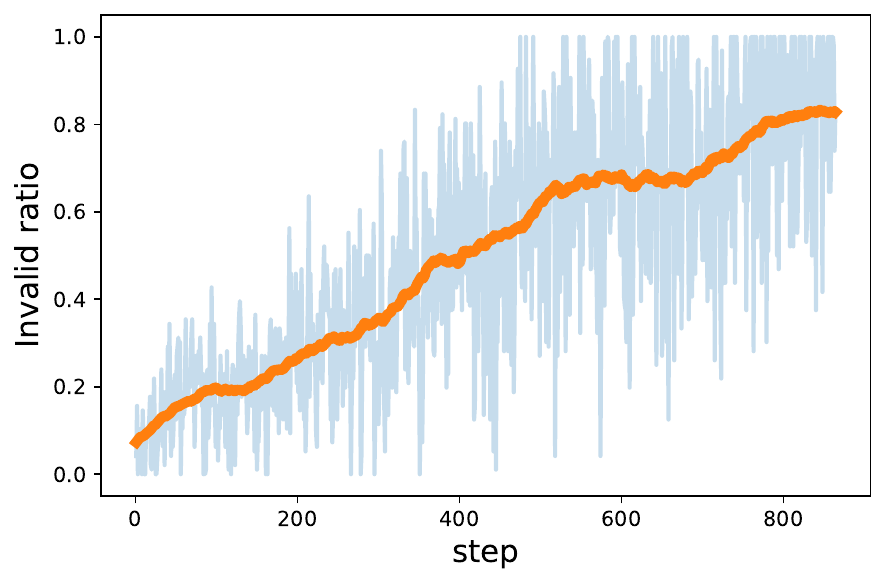}
  \caption{\textbf{Invalid proposal rate over search iterations.}}
  \label{fig:invalid_ratio}
\end{figure}

\begin{figure}[h]
  \centering
  \includegraphics[width=0.8\columnwidth]{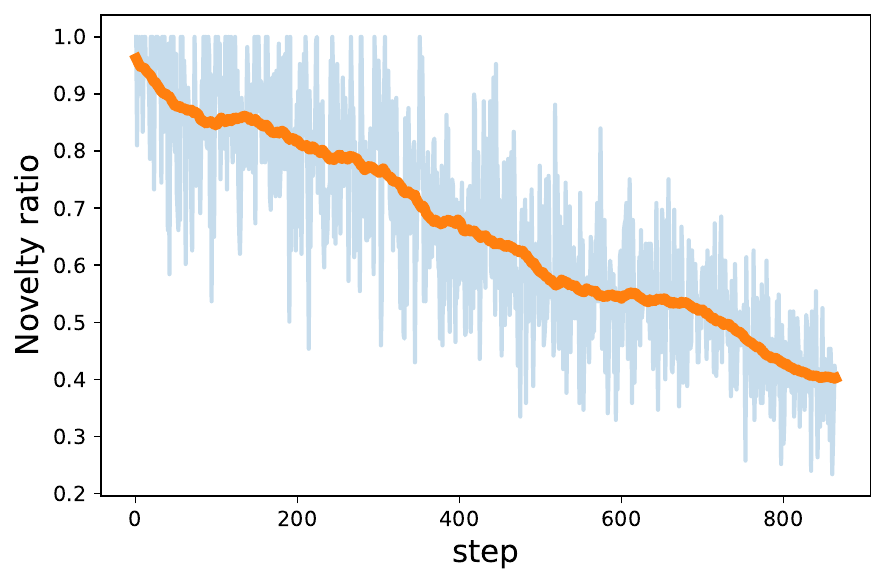}
  \caption{\textbf{Novelty acceptance rate over search iterations.}}
  \label{fig:novelty_ratio}
\end{figure}

Understanding the evolution of novelty and validity metrics during search is essential for analyzing both dataset quality and sampling efficiency. In our runs, the search process did not terminate due to full saturation of the design space, but rather due to a practical constraint: under strict validation rules, the rate of invalid proposals increases rapidly in later iterations, reaching up to $\sim$85\%. At the same time, the acceptance rate of novel samples drops to 40--50\%, indicating diminishing returns under a fixed API or compute budget. These trends are visualized in Fig.~\ref{fig:invalid_ratio} and Fig.~\ref{fig:novelty_ratio}. The former shows the steadily increasing invalidity rate as the search progresses, while the latter highlights the decreasing share of accepted novel samples. Together, they illustrate a key trade-off: although exploration can continue, its efficiency degrades substantially. Extending the process further would likely require stronger proposal strategies rather than simply running longer.

\section{Rotational augmentation}
\label{sec:rot_aug}


We apply rotational augmentation to make training robust to global orientation. In practice, the same CAD part can be stored or observed under arbitrary rotations, while its construction logic and parameterization remain unchanged. Without augmentation, the model may implicitly rely on dataset-specific canonical poses, which reduces generalization \cite{simard2003best,cohen2016steerable,weiler20183d,zhemchuzhnikov20226dcnn,zhemchuzhnikov2024ilpo,zhemchuzhnikovfourier,zhemchuzhnikov2024volumetric}. By adding randomly rotated variants of each script, we encourage the model to focus on pose-independent geometric and procedural cues and improve performance on unseen orientations. To perform rotational augmentation, we use a script-based rotation procedure. Specifically, we rotate the arguments of \textsc{CadQuery} \texttt{Workplane} construction calls, which changes the orientation and offsets of the reference planes used by subsequent operations. All remaining calls are divided into two categories: \emph{local-coordinate} and \emph{global-coordinate}. Local-coordinate functions operate in the coordinate frame of a given workplane and therefore require no changes under rotation. Global-coordinate functions, in contrast, are defined in the global frame and must be rotated accordingly. The algorithm thus detects workplane-creation and global-coordinate calls and rewrites their arguments by applying the corresponding rotation. We consider 24 rotation variants, grouped into three types:
\begin{enumerate}
    \item rotation by $0^\circ, 90^\circ, 180^\circ,$ or $270^\circ$ about the $Z$-axis;
    \item rotation by $0^\circ, 90^\circ, 180^\circ,$ or $270^\circ$ about the $Z$-axis, followed by a $90^\circ$ rotation about the $Y$-axis, and then by $0^\circ, 90^\circ, 180^\circ,$ or $270^\circ$ about the $Z$-axis;
    \item rotation by $0^\circ, 90^\circ, 180^\circ,$ or $270^\circ$ about the $Z$-axis, followed by a $180^\circ$ rotation about the $Y$-axis.
\end{enumerate}
For each dataset element, one random rotation was applied and the resulting sample was added to the training set.





\section{Comparison to existing benchmarks}
\label{sec:suppl_benchmarks}

\begin{figure}[h]
  \centering
  \includegraphics[width=0.75\columnwidth]{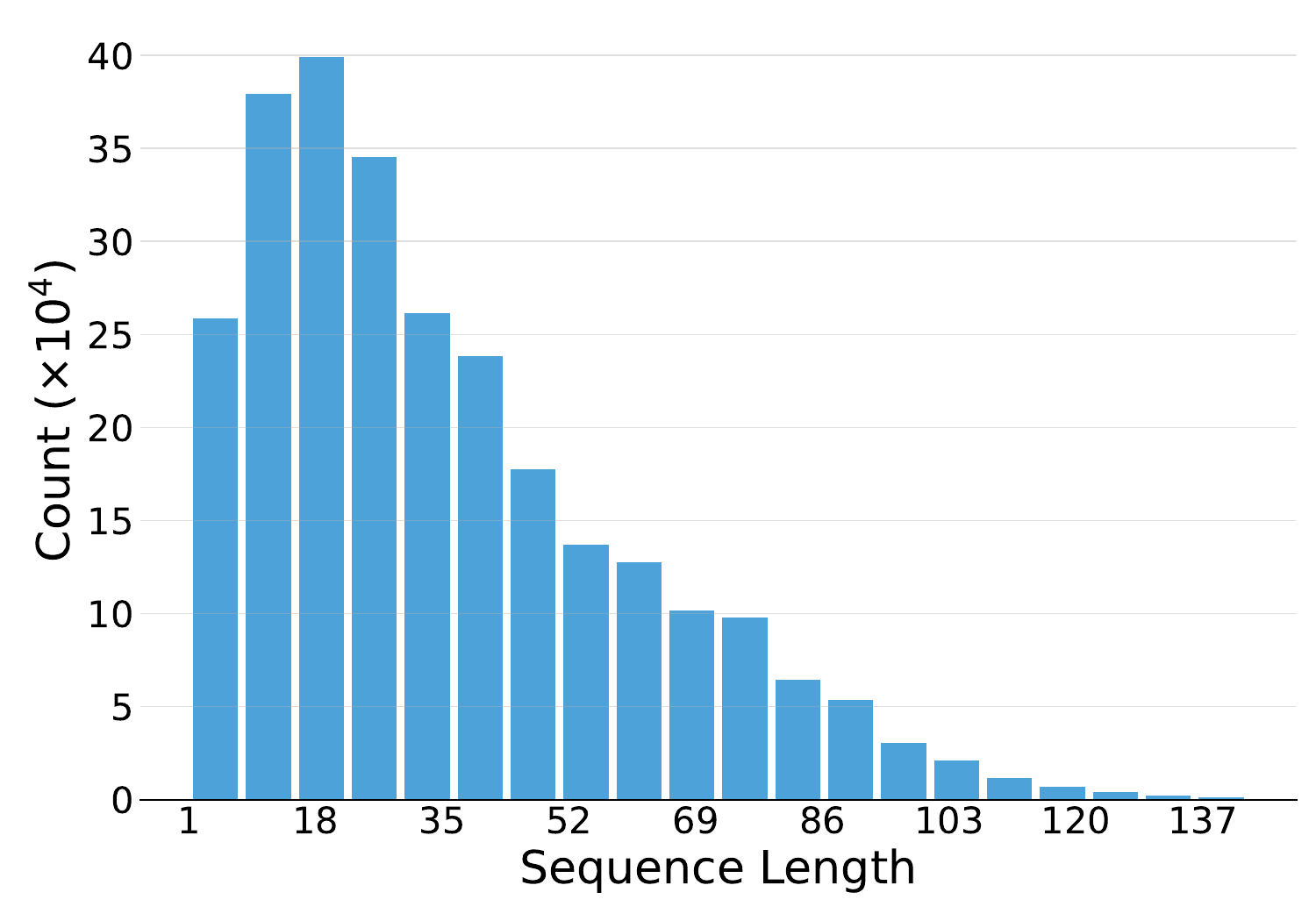}
  \caption{\textbf{Sequence length distribution.} Many CADEvolve scripts exceed 25 operations, with a long tail of highly procedural models.}
  \label{fig:seq_length_compare}
\end{figure}

\begin{figure}[h]
  \centering
  \includegraphics[width=0.75\columnwidth]{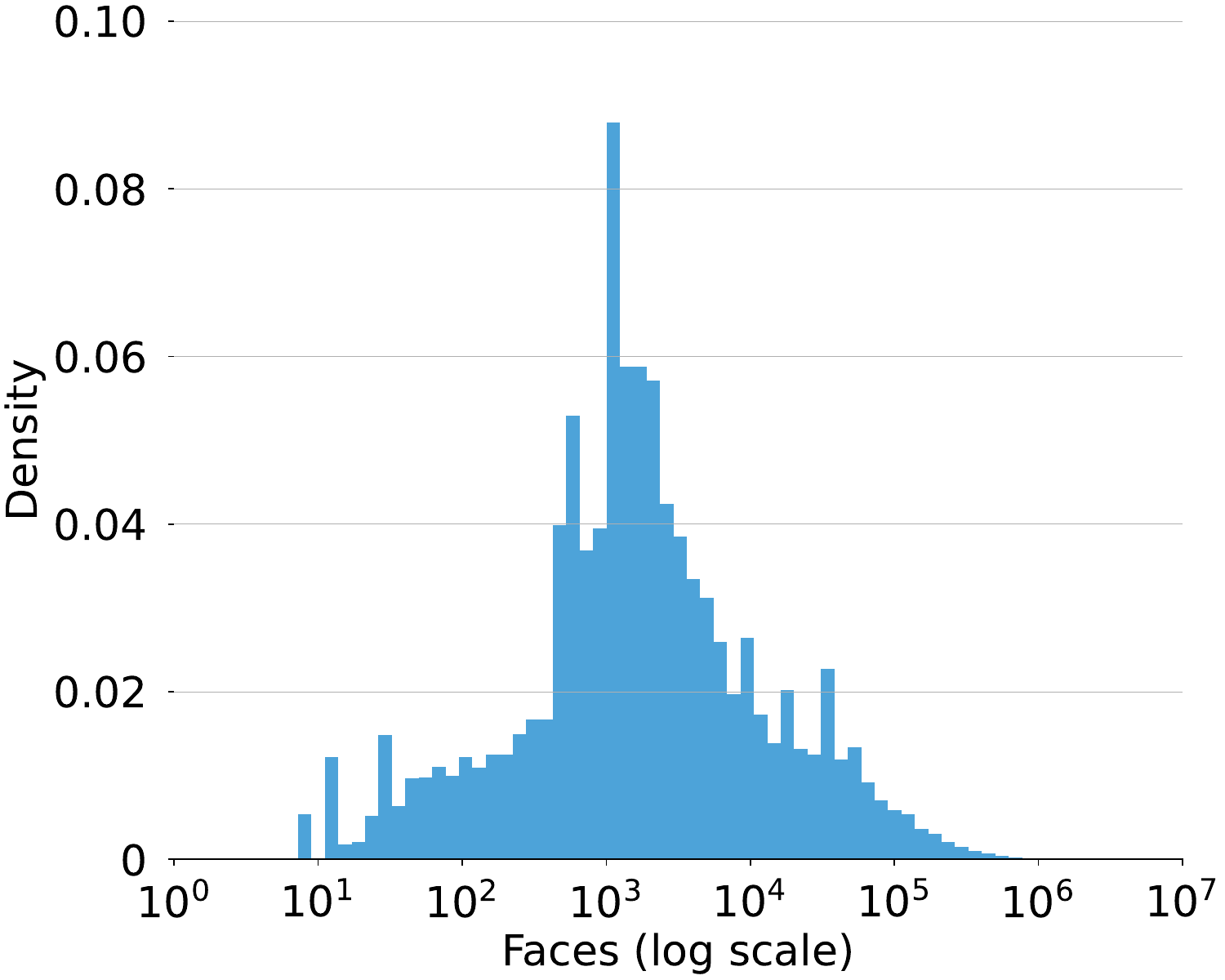}
  \caption{\textbf{Face count distribution.} CADEvolve parts frequently contain thousands of faces, reflecting fine-grained and detailed geometry.}
  \label{fig:face_count_compare}
\end{figure}

To characterize CADEvolve and compare it to existing CAD benchmarks, we report three complementary statistics: 
(i) \emph{operation occurrence} (Table~\ref{tab:operation_stats}), 
(ii) \emph{sequence length}—the number of CAD operations per script (Fig.~\ref{fig:seq_length_compare}), and 
(iii) \emph{face count}—the number of polygonal faces in the resulting geometry (Fig.~\ref{fig:face_count_compare}).
Together, these metrics capture operator coverage, procedural depth, and geometric complexity.

\paragraph{Operation occurrence statistics.}
Table~\ref{tab:operation_stats} reports the fraction of scripts that contain each \textsc{CadQuery} operation. Overall, the distribution broadly follows that of real CAD program histories, with two notable shifts: 
(i) fewer \texttt{revolve}, \texttt{chamfer}, \texttt{shell}, and \texttt{mirror} operations; and 
(ii) more \texttt{hole} operations and substantially more \texttt{transform} and \texttt{loft} operations. 
Despite these differences, the most frequent operators are present in sufficient quantities for reliable training and evaluation.

\begin{table}[h]
\centering
\caption{Operation statistics.}
\label{tab:operation_stats}
\begin{tabular}{lr}
\hline
Operation & \% \\
\hline
\texttt{extrude} & 83.05\% \\
\texttt{fillet} & 27.78\% \\
\texttt{revolve} & 4.80\% \\
\texttt{chamfer} & 4.76\% \\
\texttt{hole} & 11.99\% \\
\texttt{shell} & 1.95\% \\
\texttt{mirror} & 0.08\% \\
\texttt{sweep} & 5.75\% \\
\texttt{transform} & 20.45\% \\
\texttt{loft} & 8.48\% \\
\hline
\end{tabular}
\end{table}

\FloatBarrier

\paragraph{Sequence length.}
As shown in Fig.~\ref{fig:seq_length_compare}, CADEvolve exhibits a wide distribution of program lengths with a long tail of highly procedural models, indicating substantially greater procedural depth than typical benchmarks.

\FloatBarrier

\paragraph{Face count.}
Fig.~\ref{fig:face_count_compare} shows that CADEvolve parts frequently contain thousands of polygonal faces, reflecting fine-grained geometric detail and higher shape complexity than existing benchmarks.

\FloatBarrier


\end{document}